%% file: Main.tex
\documentclass[twocolumn,9pt]{IEEEtran}
\usepackage{times}
\usepackage[final]{graphicx}
\usepackage{amsmath,amsfonts}
\usepackage{amssymb,amsbsy}
\usepackage{times}

\usepackage{epsf}

\newtheorem{prp}{Proposition}
\newtheorem{thm}{Theorem}

\input{capscale_def}

\begin{document}
\renewcommand{\textfraction}{0}

\title{Non-Coherent Capacity and Reliability of \\ Sparse Multipath Channels in the Wideband Regime}
\author{ \emph{{Gautham Hariharan and Akbar M. Sayeed}} \\
Department of Electrical and Computer Engineering \\ University of
Wisconsin-Madison \\ \emph{gauthamh@cae.wisc.edu,
akbar@engr.wisc.edu}
\thanks{This work was partly supported by the NSF under grant \#CCF-0431088}}

\date{}
\maketitle

\input{abstract.tex}

\input{intro.tex}

\input{sysmodel.tex}

\input{ergcap.tex}

\input{reliability.tex}

\bibliographystyle{IEEEbib}
\bibliography{bib_ITAworkshop}

\end{document}

%% file: capscale_def.tex
\newcommand\ignore[1]{}

\newcommand{\hsp}{\hspace{0.05in} }

\newcommand{\bH}{\mathbf{H}  }
\newcommand{\bQ}{\mathbf{Q}}
\newcommand{\bEe}{{\mathbf{E}}}

\newcommand{\by}  { {\mathbf{y}}  }
\newcommand{\br}  { {\mathbf{r}}  }

\newcommand{\bx}  { {\mathbf{x}}  }

\newcommand{\bw} { {\mathbf{w}} }
\newcommand{\bI} { {\mathbf{I}} }

\newcommand{\snr}{{\sf{SNR}} }

\newcommand{\mseq}{{\sf{MSE}} }

\newcommand{\ord}{{\mathcal{O}}}
\newcommand{\littleo}{{\mathnormal{o}}}

\newcommand{\sebnomin} {\mathnormal{  {\frac{E_b } {N_0}_{ {\mathrm{min}} }}  } }

\newcommand{\pe} {{P}_{e}}

\newcommand{\er} {{E}_{r}}
\newcommand{\eo} {{E}_{o}}
\newcommand{\rcr} {{R}_{cr}}
\newcommand{\rmax} {{R}_{max}}

\newcommand{\Tcoh}{\mathnormal{T_{coh}}}
\newcommand{\Wcoh}{\mathnormal{W_{coh}}}
\newcommand{\Ncoh}{\mathnormal{N_{c}}}

\newcommand{\nc}{\mathnormal{N_{c}}}

\newcommand{\eff}{{\mathrm{eff}} }

\newcommand{\Dwmax} { D_{W,\max} }
\newcommand{\Dtmax} { D_{T,\max} }
\newcommand{\nsnr} { N_{c} \snr }
\newcommand{\etas} {\eta^{*}}

\newcommand{\tsty} {\textstyle}

%% file: abstract.tex
\begin{abstract}
In contrast to the prevalent assumption of rich multipath in
information theoretic analysis of wireless channels, physical
channels exhibit sparse multipath, especially at large bandwidths.
We propose a model for sparse multipath fading channels and
present results on the impact of sparsity on non-coherent capacity
and reliability in the wideband regime. A key implication of
sparsity is that the statistically independent degrees of freedom
in the channel, that represent the delay-Doppler diversity
afforded by multipath, scale at a {\em sub-linear} rate with the
signal space dimension (time-bandwidth product). Our analysis is
based on a training-based communication scheme that uses
short-time Fourier (STF) signaling waveforms. Sparsity in
delay-Doppler manifests itself as time-frequency coherence in the
STF domain. From a capacity perspective, sparse channels are
asymptotically coherent: the gap between coherent and non-coherent
extremes vanishes in the limit of large signal space dimension
without the need for peaky signaling. From a reliability
viewpoint, there is a fundamental tradeoff between channel
diversity and learnability that can be optimized to maximize the
error exponent at any rate by appropriately choosing the signaling
duration as a function of bandwidth.

\end{abstract}

%% file: intro.tex
\section{Introduction}

Recent advances in the emerging areas of ultra-wideband
communication systems and wireless sensor networks have renewed
the search for a complete understanding of the fundamental
performance limits in the wideband/low $\snr$ regime. The impact
of multipath signal propagation, which leads to fading, on the
capacity and reliability of wideband channels, depends critically
on knowledge of the channel state information (CSI) at the
receiver. The seminal work in \cite{verdu} best illustrates this:
with perfect CSI at the receiver, peak-power limited QPSK achieves
second-order optimality, whereas with no receiver CSI, peaky
signals are necessary to even achieve first-order optimality,
although they fail to be second-order optimal. Motivated by the
fact that channel learning can bridge this gap and the sharp
cut-off, the authors in \cite{zheng} study wideband capacity by
assuming a coherence time scaling with $\snr$ of the form
\begin{equation}
\Tcoh = \frac{k}{\snr^{\mu}} \hsp , \hsp \hsp \mu > 0
\end{equation}
However there is no explanation for why such scaling laws should
hold in practice.

Accurate modeling of the channel characteristics in time and
frequency, as a function of physical multipath characteristics, is
critical in analyzing the performance of channel learning schemes
and the impact of CSI on the performance limits. While most
existing results assume rich multipath, there is growing
experimental evidence (e.g. \cite{molisch, saadane}) that physical
channels exhibit a sparse structure at wide bandwidths and when we
code over long signaling durations. In this paper, we use a
virtual representation \cite{akbar_behnaam} for physical multipath
channels to present a framework for modeling sparsity. The virtual
representation uniformly samples multipath in delay and Doppler at
a resolution commensurate with the signaling bandwidth and
signaling duration, respectively. Under this representation, the
virtual channel coefficients represent the DoF in delay and
Doppler. Sparse channels correspond to a sparse set of virtual
coefficients and a key implication is the sub-linear scaling of
the number of independent degrees of freedom (DoF) with signal
space dimensions. This is contrast to rich multipath, where the
DoF scale linearly. We consider signaling over orthogonal
short-time Fourier (STF) basis functions that serve as approximate
eigenfunctions for underspread channels and provide a natural
mechanism to relate sparsity in delay-Doppler to coherence in
time-frequency.

With no receiver CSI \emph{a priori}, we consider training-based
communication schemes and investigate the wideband ergodic
capacity under the assumption that the \emph{time-frequency
coherence dimension} $\nc$ scales with $\snr$ according to
\begin{equation}
\nc = \frac{k}{\snr^{\mu}} \hsp , \hsp \hsp \mu > 0
\label{ncsnr_intro}
\end{equation}
It is observed that the coherence requirements for achieving
capacity are shared between both time and frequency: the coherence
bandwidth, $W_{coh}$, increases with bandwidth, $W$ (due to
sparsity in delay), and the coherence time, $T_{coh}$, increases
with signaling duration $T$ (due to sparsity in Doppler). As a
result, the scaling requirements on $T_{coh}$ with $W$ needed in
\cite{zheng} for first- and second-order optimality are replaced
by scaling requirements on $N_{c} = T_{coh} W_{coh}$. This leads
to dramatically relaxed requirements on $T_{coh}$ scaling with
bandwidth/SNR compared to those assumed in \cite{zheng}. In
particular, sparse multipath channels are {\em asymptotically
coherent}; that is, for a sufficiently large but fixed bandwidth,
the conditions for first- and second-order optimality can be
achieved by simply making the signaling duration sufficiently
large according to
\begin{equation}
T \propto \frac{\left( T_{m}^{\delta_2} W_{d}^{\delta_1}
\right)^{\frac{1}{1-\delta_1}} W^{\frac{\mu - 1 +
\delta_2}{1-\delta_1}}}{P^{\frac{\mu}{1-\delta_1}}} \label{TWP}
\end{equation}
Equation (\ref{TWP}) relates the signaling parameters
($T$,$W$,$P$), as a function of the channel parameters
($T_{m}$,$W_{d}$,$\delta_1$,$\delta_2$) in order for the
relationship (\ref{ncsnr_intro}) to hold between $\nc$ and $\snr$
at any desired value of $\mu$ (in particular, $\mu>1$ for
first-order optimality and $\mu>3$ for second-order optimality).
The asymptotic coherence of sparse channels also eliminates the
need for peaky signaling that has been emphasized in existing
results \cite{kennedy,verdu} for increasing the spectral
efficiency of non-coherent communication schemes.

Our investigation of the reliability of sparse channels is through
random coding error exponents \cite{gallager_book}. For
training-based communication schemes, our results reveal a
fundamental \emph{learnability versus diversity tradeoff} in
sparse channels. At any transmission rate less than the coherent
capacity, there is an optimal choice of signal parameters (as a
function of channel parameters) that optimizes the tradeoff and
yields the largest error exponent.

%% file: sysmodel.tex
\section{System Setup}
\label{sec2}
\subsection{Sparse Multipath Channel Modeling}
\label{sec2a} 
\begin{figure}[h]
\begin{center}
\begin{tabular}{cc}
\begin{minipage}{1.6in}
\centerline{\includegraphics[width=1.6in]{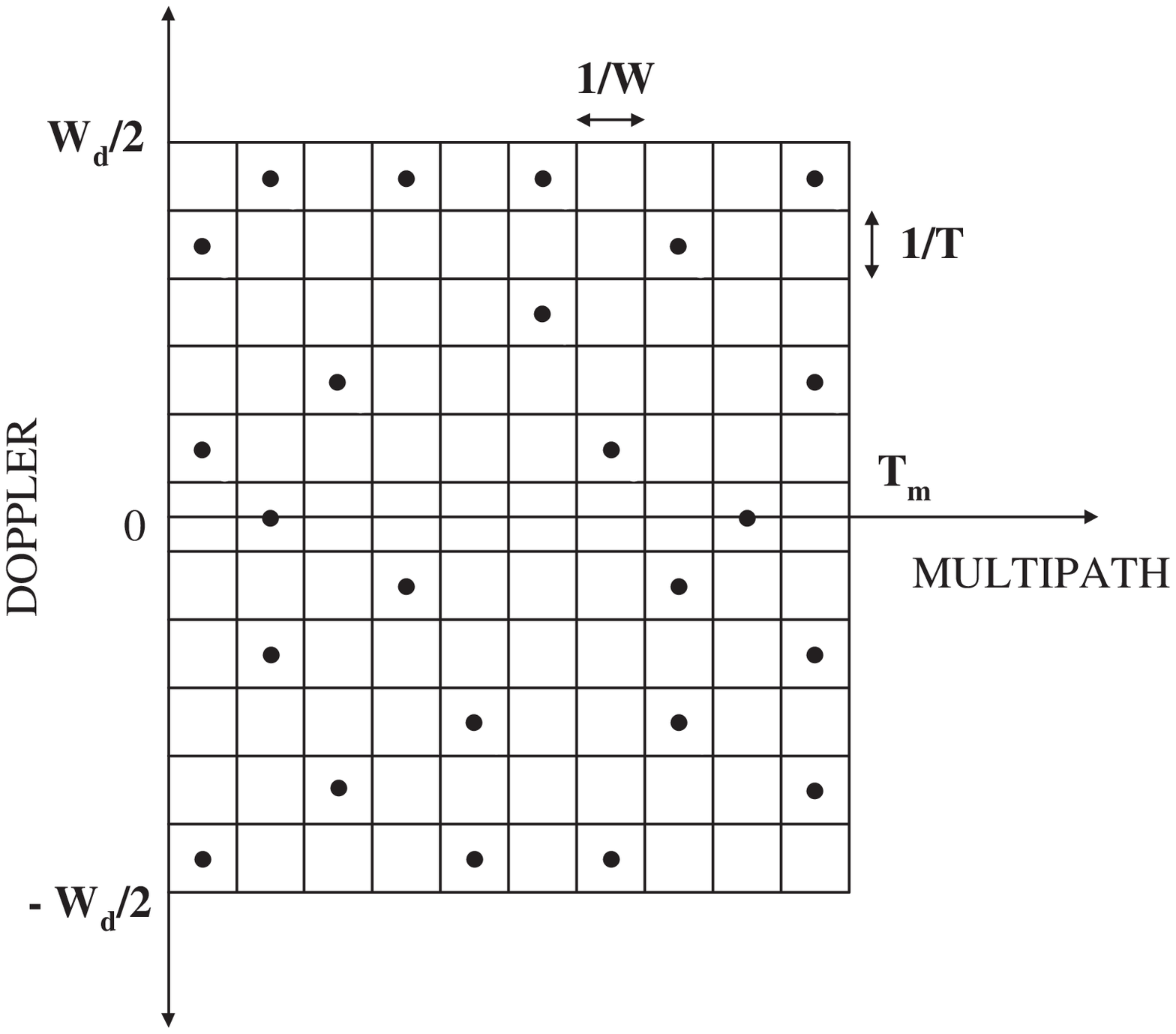}}
\end{minipage} &
\begin{minipage}{1.9in}
\centerline{\includegraphics[width=1.9in]{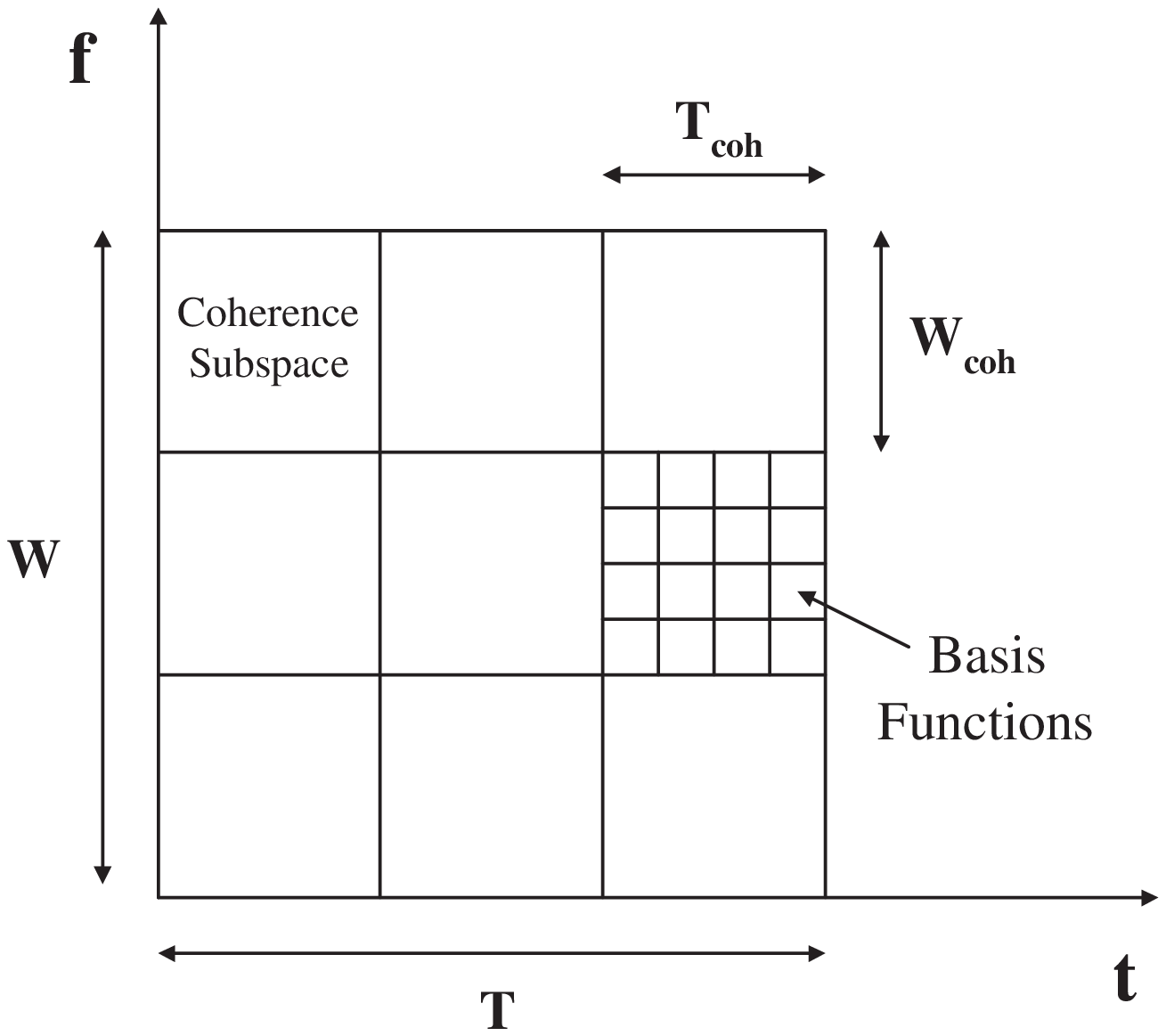}}
\end{minipage} \\
(a) & (b)
\end{tabular}
\includegraphics[width=2.0in]{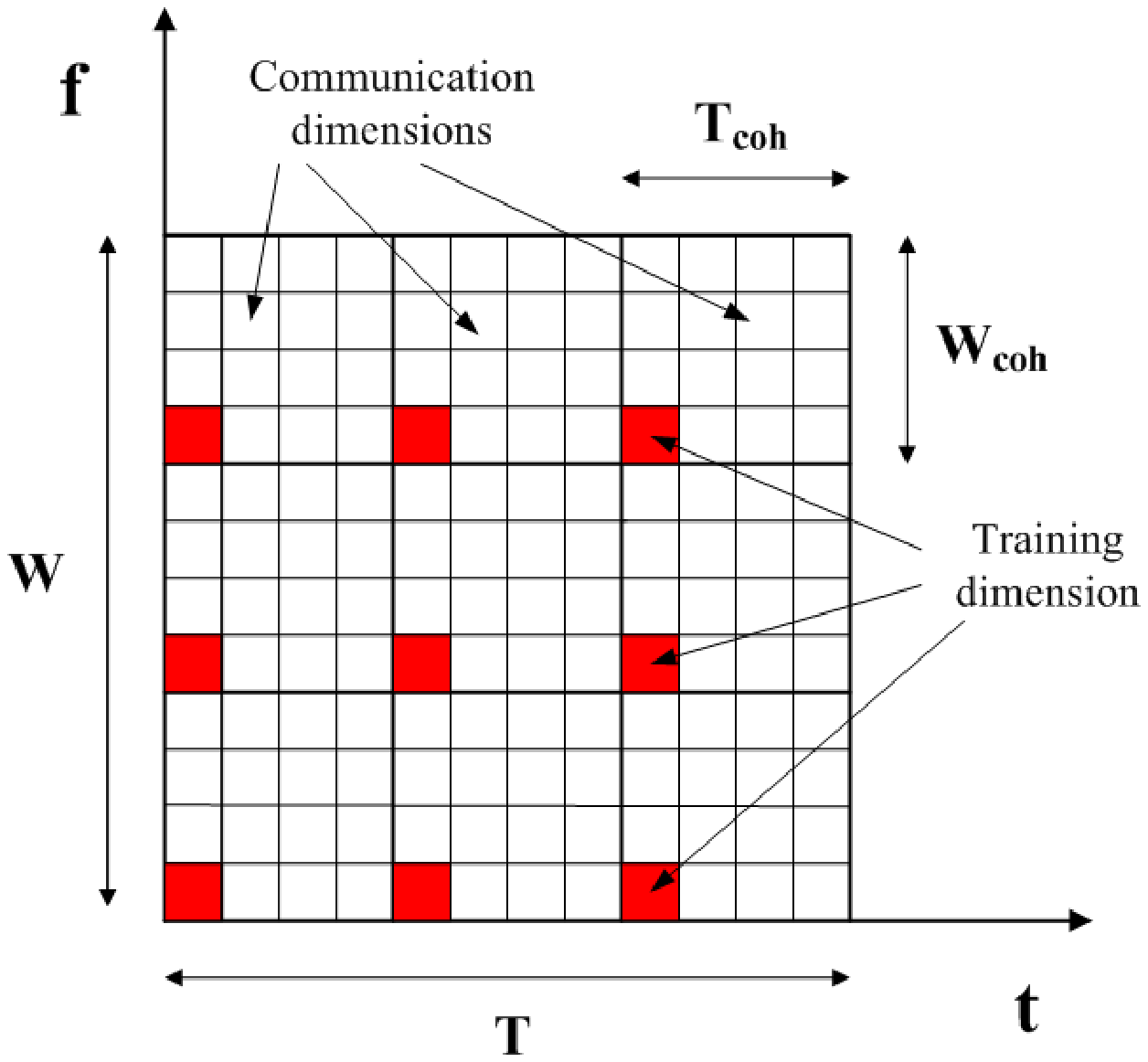}\\
(c) \caption{ \label{fig:del_dopp} {\sl (a) Delay-doppler sampling
commensurate with signaling duration and bandwidth. (b)
Time-frequency coherence subspaces in STF signaling. (c)
Illustration of the training-based communication scheme in the STF
domain. One dimension in each coherence subspace (dark squares)
represent the training dimension and the remaining dimensions are
used for communication.}}
\end{center}
%\vspace{-5mm}
\end{figure}

A physical discrete multipath channel can be modeled
as
\begin{eqnarray}
h(\tau,\nu) & = & \sum_{n} \beta_n \delta(\tau-\tau_n) \delta(\nu
- \nu_n) \nonumber \\
r(t) & = & \sum_n \beta_n x(t-\tau_n)e^{j2\pi \nu_n t} + w(t)
\label{disc_mp}
\end{eqnarray}
where $h(\tau,\nu)$ is the delay-Doppler spreading function of the
channel, $\beta_n$, $\tau_n \in [0,T_m]$ and $\nu_n \in
[-W_d/2,W_d/2]$ denote the complex path gain, delay and Doppler
shift associated with the $n$-th path. $T_{m}$ and $W_{d}$ are the
delay and Doppler spreads respectively and $w(t)$ is additive
white Gaussian noise (AWGN). We assume a sufficiently underspread
channel, $T_{m}W_{d} \ll 1$. In this paper we use a \emph{virtual
representation} \cite{akbar_behnaam, akbar_and_venu} for time- and
frequency-selective multipath channels that captures the channel
characteristics in terms of {\em resolvable paths} and greatly
facilitates system analysis from a communication-theoretic
perspective. The virtual representation uniformly samples the
multipath in delay and Doppler at a resolution commensurate with
signaling bandwidth $W$ and signaling duration $T$, respectively
\cite{akbar_behnaam,akbar_and_venu}
\begin{eqnarray}
y(t) &=& \sum_{\ell=0}^{\lceil T_m W\rceil} \sum_{m=-\lceil T
W_d/2\rceil}^{\lceil TW_d/2\rceil}h_{\ell,m} x(t- \ell/W)e^{j2\pi
mt/T}  \label{del_dopp_samp} \\
\ h_{\ell,m} &\approx&  \sum_{n \in S_{\tau,\ell} \cap S_{\nu,m}}
\beta_n  \label{hlm}
\end{eqnarray}
The sampled representation (\ref{del_dopp_samp}) is linear and is
characterized by the virtual delay-Doppler channel coefficients
$\{ h_{\ell,m} \}$. Each $h_{\ell,m}$ consists of the sum of gains
of all paths whose delays and Doppler shifts lie within the
$(\ell,m)$-th delay-Doppler resolution bin as shown in
Fig.~\ref{fig:del_dopp}(a). Distinct $h_{\ell,m}$'s correspond to
approximately \emph{disjoint} subsets of paths and are hence
approximately statistically independent (due to independent path
gains and phases). In this work, we assume that the channel
coefficients $\{ h_{\ell,m}\}$ are perfectly independent. We also
assume Rayleigh fading in which $\{ h_{\ell,m}\}$ are zero-mean
Gaussian random variables and the channel statistics are thus
characterized by the power in the virtual channel coefficients
\begin{eqnarray}
\Psi(\ell,m) &=& E[|h_{\ell,m}|^2] \label{hlm_power}
\end{eqnarray}
which is a measure of the (sampled) delay-Doppler power spectrum.

Let $D$ denote the number of dominant \footnote{For which
$\Psi(\ell,m) > \gamma$ for some prescribed threshold $\gamma >
0$.} non-zero channel coefficients. The parameter $D$ reflects the
statistically independent degrees of freedom (DoF) in the channel
and also signifies the delay-Doppler diversity afforded by the
channel. It can be bounded as
\begin{eqnarray}
D  =  D_{T} D_{W} \leq D_{\max} = \Dtmax \Dwmax \nonumber \\
\Dtmax = \left \lceil TW_d  \right \rceil \ , \ \Dwmax =
\left\lceil T_m W \right \rceil  \label{del_dopp_div}
\end{eqnarray}
where $\Dtmax$ denotes the maximum number of resolvable paths in
Doppler (maximum Doppler or time diversity) and $\Dwmax$ denotes
maximum number of resolvable paths in delay (maximum delay or
frequency diversity). In rich multipath, $D_{T} = \Dtmax$ and
$D_{W} = \Dwmax$ and each delay-Doppler resolution bin in
Fig.~\ref{fig:del_dopp}(a) is populated by a path. In this case
$D$ scales linearly with the signal space dimensions, $N = TW$.

However, recent measurement campaigns~\cite{molisch,saadane} for
UWB channels show that dominant channel coefficients get sparser
in the delay domain as the bandwidth increases. As we consider
large bandwidths and/or long signaling durations, the resolution
of paths in both delay and Doppler domains gets finer, leading to
the scenario in Fig.~\ref{fig:del_dopp}(a) where the delay-Doppler
resolution bins are sparsely populated with paths, i.e. $D <
D_{max}$. Thus physical multipath channels get sparser with
increasing $W$ due to fewer than $\Dwmax$ resolvable delays and
with increasing $T$ due to fewer than $\Dtmax$ resolvable Doppler
shifts. We model such sparse behavior with a {\em sub-linear}
scaling in $D_T$ and $D_W$ with $T$ and $W$:
\begin{equation}
D_T \sim  (TW_d)^{\delta_1} \ , \ D_W \sim (T_mW)^{\delta_2} \ , \
\delta_1, \delta_2 \in [0,1] \label{sparse}
\end{equation}
where $\{ \delta_{i} \}$ represent channel sparsity; smaller the
value of $\{ \delta_i \}$, the slower (sparser) the growth in the
resolvable paths in the corresponding domain. This implies that
the delay-Doppler DoF, $D = D_T D_W$, scale sub-linearly with the
number of signal space dimensions $N$. Note that with perfect CSI
at the receiver, $D$ reflects the delay-Doppler diversity afforded
by the channel, whereas with no CSI, it reflects channel
uncertainty.

\subsection{Orthogonal Short-Time Fourier Signaling}
\label{sec2b} We consider signaling using an orthonormal
short-time Fourier (STF) basis~\cite{ke_tamer_say,kozek} that is a
natural generalization of orthogonal frequency-division
multiplexing (OFDM) for time-varying channels. An orthogonal STF
basis for the signal space is generated from a fixed prototype
waveform $g(t)$ via time and frequency shifts:
\begin{eqnarray}
\phi_{\ell,m}(t) = g(t-\ell T_o)e^{j2\pi W_ot} \hsp \hsp
\text{where} \hsp \hsp T_o W_o = 1 \nonumber \\
\ell = 0, \hdots ,N_{T}-1, \hsp \hsp m = 0, \hdots, N_{W}-1
\nonumber \\
N_{T} = \frac{T}{T_o}, N_{W} = \frac{W}{W_o} \hsp \text{and} \hsp
N = N_{T}N_{W}
\end{eqnarray}
The $N$ transmitted symbols $x_{\ell,m}$ are modulated onto the
STF basis
\begin{equation*}
x(t) = \sum \limits_{\ell,m} x_{\ell,m}\phi_{\ell,m}(t)
\end{equation*}
For a signaling duration $T$ and bandwidth $W$, the basis
functions span the signal space with dimension equal to $N=TW$.

The received signal is given by
\begin{equation*}
r(t) = \mathcal{H}\left(x(t)\right) + w(t)
\end{equation*}
The received signal is projected onto the STF basis waveforms to
yield the received symbols
\begin{equation}
r_{\ell,m} = \langle r, \phi_{\ell,m} \rangle = \sum
\limits_{\ell^{'},m^{'}} h_{\ell,m \;\ \ell^{'},\; m^{'}} \;\
x_{\ell^{'},\; m^{'}} + w_{\ell,m} \label{stf_rx}
\end{equation}
Equivalently, we can represent the system in STF-domain using an
$N$-dimensional matrix system equation
\begin{equation}
\br  =  \sqrt{\snr} \hsp {\mathbf{H}} \bx + \bw
\label{disc_channel}
\end{equation}
where $\bw$ represents the additive noise vector whose entries are
i.i.d.\ $\mathrm{CN}(0,1)$. The $N \times N$ matrix consists of
the channel coefficients $\{h_{\ell,m \;\ \ell^{'},\; m^{'}}\}$ in
(\ref{stf_rx}). The parameter $\snr$ represents the transmit
energy per modulated symbol and for a given transmit power $P$
equals $\snr = \frac{P}{W}$ ($\bEe[|\bx|^2]=1$). In this work, our
focus is on the wideband regime, where $\snr \rightarrow 0$.

For sufficiently underspread channels, the parameters $T_o$ and
$W_o$ can be matched to $T_m$ and $W_d$ so that the STF basis
waveforms serve as approximate eigenfunctions of the channel
\cite{kozek,ke_tamer_say}. Thus the $N \times N$ channel matrix
$\bH$ is approximately diagonal. In this work, we will assume that
$\bH$ is exactly diagonal, that is,
\begin{equation}
{\mathbf{H}}  =  {\mathrm{diag}} \Big[ \underbrace{ {h}_{1,1}
\cdots {h}_{1,\Ncoh}}_{  {\mathrm{Subspace}} \hsp 1}, \hsp
\underbrace { {h}_{2,1} \cdots {h}_{2,\Ncoh} }_{
{\mathrm{Subspace}} \hsp 2} \hsp  \cdots \hsp \underbrace {
{h}_{D,1} \cdots {h}_{D,\Ncoh} }_{ {\mathrm{Subspace}} \hsp D}
\Big] \ .\label{H_diag}
\end{equation}
Furthermore, the diagonal entries of $\bH$ in (\ref{H_diag}) admit
an intuitive block fading interpretation in terms of {\em
time-frequency coherence subspaces} \cite{ke_tamer_say}
illustrated in Fig.~\ref{fig:del_dopp}(b). The signal space is
partitioned as
\begin{equation}
N = TW=\nc D \label{N_DNc} \end{equation} where $D$ represents the
number of statistically independent time-frequency coherence
subspaces (delay-Doppler diversity), reflecting the DoF in the
channel (see (\ref{del_dopp_div})), and $\nc$ represents the
dimension of each coherence subspace, which we will refer to as
the \textbf{time-frequency coherence dimension}. In the block
fading model in (\ref{H_diag}), the channel coefficients over the
$i$-th coherence subspace $h_{i,1} \cdots h_{i,\Ncoh}$ are assumed
to be identical, $h_i$, whereas the coefficients across different
coherence subspaces are independent and due to the stationarity of
the channel statistics across time and frequency, identically
distributed. Thus, the $D$ distinct STF channel coefficients, $\{
h_i \}$, are i.i.d. zero-mean Gaussian random variables (Rayleigh
fading) with variance $\bEe[|h_i|^2]=1$.

Using the DoF scaling for sparse channels in (\ref{sparse}), the
coherence dimension of each coherence subspace can be computed as
\begin{eqnarray}
\Tcoh = \frac{T}{D_T} = \frac{T^{1-\delta_1}}{W_d^{\delta_1}} \ ,
\ \Wcoh = \frac{W}{D_W} = \frac{W^{1-\delta_2}}{T_m^{\delta_2}}
\label{Wcoh} \\
 \nc = \Tcoh \Wcoh
  = \frac{T^{1-\delta_1}}{W_d^{\delta_1}}
\frac{W^{1-\delta_2}}{T_m^{\delta_2}} \geq \left \lceil
\frac{1}{T_mW_d} \right \rceil = \nc_{,\min} \label{Ncoh}
\end{eqnarray}
where $\Tcoh$ is  the {\em coherence time} and $\Wcoh$ is the {\em
coherence bandwidth} of the channel, as illustrated in
Fig.~\ref{fig:del_dopp}(b). Note that $\delta_1 = \delta_2 = 1$
corresponds to a rich multipath channel in which $\nc =
\nc_{,\min} = 1/(T_m W_d)$ is \emph{constant} and $D = D_{\max}$
increases \emph{linearly} with $N = TW$. This is the assumption
prevalent in existing works. In contrast, for sparse channels,
$(\delta_1, \delta_2) \in (0,1)$, and both $\nc$ and $D$ increase
{\emph{sub-linearly}} with $N$. In terms of channel parameters,
$\nc$ increases with decreasing $T_{m}W_{d}$ as well as with
smaller $\delta_{i}$. In terms of signaling parameters, $\nc$ can
be increased by increasing $T$ and/or $W$. On the other hand, when
the channel is rich, $\nc$ depends only on $T_{m}W_{d}$ and does
not scale with $T$ or $W$.

In this paper, our focus is on computing the sparse channel
capacity and reliability and, as we will see later, both metrics
turn out to be functions only of the parameters $\nc$ and $\snr$.
Furthermore, in the wideband limit they critically depend on the
following relation between $\nc$ and $\snr$
\begin{equation}
\nc  =  \frac{k}{\snr^{\mu}} \hsp , \hsp \hsp \hsp \mu > 0
\label{nc_snr_relation}
\end{equation}
where $k>0$ is a constant.

\subsection{Training-Based Communication Using STF Signaling}
\label{sec2c} We use the block fading model induced by STF
signaling to study the impact of time-frequency coherence on
channel capacity and reliability in sparse multipath channels.
Within the non-coherent regime, we focus our attention on a
communication scheme in which the transmitted signals include
training symbols to enable coherent detection. Although it is
argued in \cite{zheng} that training-based schemes are sub-optimal
from a capacity point of view, the restriction to training schemes
is motivated by practical considerations.

We provide an outline of the training-based communication scheme,
adapted from \cite{zheng}, suitable to STF signaling (see
\cite{capacity_sparse_STSP} for details). The total energy
available for training and communication is $PT$, of which a
fraction $\eta$ is used for training and the remaining fraction
$(1-\eta)$ is used for communication. Since the quality of the
channel estimate over one coherence subspace depends only on the
training energy and {\emph{not}} on the number of training
symbols~\cite{hassibi_training}, our scheme uses one signal space
dimension in each coherence subspace for training and the
remaining $\left( \nc-1 \right)$ for communication, as illustrated
in Fig.~\ref{fig:del_dopp}(c). We consider minimum mean squared
error (MMSE) estimation under which the channel estimation
performance is measured in terms of the resulting mean squared
error ($\mseq$).

%% file: ergcap.tex
\section{Ergodic Capacity of the Training-Based Communication Scheme}
%{\vspace{-3mm}}
\label{sec3} We first characterize the coherent capacity of the
wideband channel with perfect CSI at the receiver. The coherent
capacity per dimension (in bps/Hz) is defined as
\begin{equation*}
C_{coh} \left( \snr \right) = \sup_{ {\bQ}: \hsp
{\mathrm{Tr}}({\bQ}) \hsp \leq \hsp TP  } \frac{ {\bEe} \left[
\log_2 \det \left( \bI_{N} + {\bH} {\bQ} {\bH}^H \right) \right]
}{N}
\end{equation*}
where $P$ denotes transmit power and $\bH$ is the diagonal channel
matrix in (\ref{H_diag}) with the diagonal elements following the
block-fading structure. Due to the diagonal nature of $\bH$, the
optimal $\bQ$ is also diagonal. In particular, uniform power
allocation $\bQ = \frac{TP} {N} \bI_{N} = \snr \hsp \bI_{N}$
achieves capacity and we have
\begin{eqnarray}
C_{coh} \left( \snr \right) & = &  \frac{ \sum_{i=1}^D
{\bEe}\left[ \log_2 \left( 1 + \frac{TP}{N} \left| h_i
\right|^2 \right) \right] }{D} \nonumber \\
& \stackrel{\mathit{(a)}}{=} &  {\bEe}\left[ \log_2 \left( 1 +
{\snr} \left| h \right| ^2 \right)  \right] \label{exact_cap}
\end{eqnarray}
where (a) follows since $\{ h_i \}$ are i.i.d.\ with $h_{\cdot}$
representing a generic random variable, $N = TW$ and $\snr =
\frac{P}{W} = \frac{TP}{N}$.

The next proposition provides a lower bound to the coherent
capacity in the low $\snr$ regime \cite{capacity_sparse_STSP}.
\begin{prp}
\label{lowerbound} The coherent capacity, $C_{coh} \left( \snr
\right)$, satisfies
\begin{eqnarray}
C_{coh} \geq \log_2(e) \left(\snr - \snr^2 \right)
\label{coh_bounds}
\end{eqnarray}
Moreover the capacity converges to the lower bound in the limit of
$\snr \rightarrow 0$.
%\endproof
\end{prp}
The lower bound in Proposition \ref{lowerbound} shows that the
minimum energy per bit necessary for reliable communication is
given by $\sebnomin = \log_e(2)$ and the wideband slope $S_0 = 1$,
the two fundamental metrics of spectral efficiency in the wideband
regime defined in \cite{verdu}.

In terms of the scaling law, $\nc = \frac{k}{\snr^{\mu}}, \mu >
0$, as defined in~(\ref{nc_snr_relation}), we are interested in
computing the value of $\mu$ such that the training-based
communication scheme achieves first- and second-order optimality.
The result is summarized in the following theorem.
\begin{thm}
{\label{thm1}} The average mutual information of the
training-based scheme, with the scaling law $\nc =
\frac{k}{\snr^{\mu}}$, satisfies
\begin{eqnarray}
I_{tr} \geq \log_2(e) \cdot \left[ \snr - \ord \left(
\snr^{\frac{1+\mu}{2}} \right) \right] \label{cap_tr}
\end{eqnarray}
In particular, the first- and second-order optimality conditions
are met if and only if $\mu > 1$ and $\mu > 3$, respectively.
\end{thm}
\begin{proof}
Omitted for this version. See \cite{capacity_sparse_STSP} for
details.
\end{proof}

It can be seen that the coherence dimension $\nc$ plays a critical
role in determining the capacity of the training and communication
scheme. Both channel and signaling parameters impact $\nc$ in
sparse channels and for a given $T_{m}W_{d}$ and $\{ \delta_i \}$,
the signal space parameters $T$ and $W$ can be suitably chosen to
obtain any desired value for $\nc$. Recalling the expression for
$\Wcoh$ in (\ref{Wcoh}), we note that
\begin{equation}
\Wcoh = \frac{W^{1-\delta_2}}{(T_{m})^{\delta_2}} =
\frac{P^{1-\delta_2}}{(T_{m})^{\delta_2} \snr^{1-\delta_2}}
\label{Wcoh_snr}
\end{equation}
and thus $\Wcoh$ \emph{naturally} scales with $\snr$. Using
(\ref{Wcoh_snr}) the expression for $\nc$ in  (\ref{Ncoh}) becomes
\begin{equation}
\nc = \frac{T^{1-\delta_1}}{(W_{d})^{\delta_1}}
\frac{P^{1-\delta_2}}{(T_{m})^{\delta_2}\snr^{1-\delta_2}}
\label{nc_wcoh}
\end{equation}

Equating  (\ref{nc_snr_relation}) with (\ref{nc_wcoh}) leads to
the following canonical relationship
\begin{equation}
T = \frac{\left( k \right)^{\frac{1}{1-\delta_1}}\left(
T_{m}^{\delta_2} W_{d}^{\delta_1} \right)^{\frac{1}{1-\delta_1}}
W^{\frac{\mu - 1 +
\delta_2}{1-\delta_1}}}{P^{\frac{\mu}{1-\delta_1}}}
\label{TWP_locus}
\end{equation}
that relates the signaling parameters ($T$,$W$,$P$), as a function
of the channel parameters ($T_{m}$,$W_{d}$,$\delta_1$,$\delta_2$)
in order for the relationship (\ref{nc_snr_relation}) to hold
between $\nc$ and $\snr = P/W$. Equations (\ref{nc_snr_relation})
and (\ref{TWP_locus}) are the two key equations that capture the
essence of the results in this paper.

\subsection{Discussion of Results on Ergodic Capacity}
\label{sec3a}

In the context of existing results in \cite{zheng} that assume
rich multipath ($\delta_1 = \delta_2 = 1$), Theorem~\ref{thm1}
shows that the requirement on $\Tcoh$ is now the requirement on
the coherence dimension $\nc = \Tcoh \Wcoh$. Thus, the coherence
cost is shared in both time and frequency resulting in
\emph{significantly weakened} scaling requirements for $\Tcoh$. If
we have $\Wcoh = \ord \left(W^{1-\delta_2} \right)$, then the
$\Tcoh$ scaling requirement reduces to
\begin{equation}
\Tcoh = \nc / \Wcoh = \ord \left(W^{2 + \delta_2} \right)
\label{tcoh_scaling}
\end{equation}
to achieve second-order optimality. This is significantly less
stringent than the $\Tcoh = \ord \left(W^{3} \right)$ required in
the framework of \cite{zheng}.

Combining Theorem~\ref{thm1} with (\ref{TWP_locus}) lead to
scaling rules for the locus of points ($T$,$W$,$P$) in order to
achieve a desired value of $\mu$ (Recall $\mu > 1$ for first-order
optimality and $\mu>3$ for second-order optimality). Specifically,
\begin{equation}
\begin{split}
\log\left( T \right) = & \frac{1}{1-\delta_1} \log \left(
W_{d}^{\delta_1} T_{m}^{\delta_2} \right) + \left( \frac{\mu +
\delta_2 - 1}{1-\delta_1} \right) \log \left( W \right) \\ &  -
\left( \frac{\mu}{1-\delta_1} \right) \log \left( P \right) \
.\label{TvsW}
\end{split}
\end{equation}
It is observed that smaller $\delta_i$'s imply a slower scaling of
$T$ with $W$. Conversely, for any system operating at a particular
$T$ and $W$, (\ref{TvsW}) can be used to determine the effective
value of $\mu$ as
\begin{equation}
\mu_{\eff} = \frac{\left(1-\delta_1 \right) \log(T/c) +
\left(1-\delta_2 \right) \log(P)}{\log(W/P)}  + \left(1-\delta_2
\right) \label{mueff}
\end{equation}
where $c = \left(T_{m}^{\delta_2} W_{d}^{\delta_1}
\right)^{\frac{1}{1-\delta_1}}$.

\begin{figure}[width=2.8in]
\begin{center}
\includegraphics[width=2.8in]{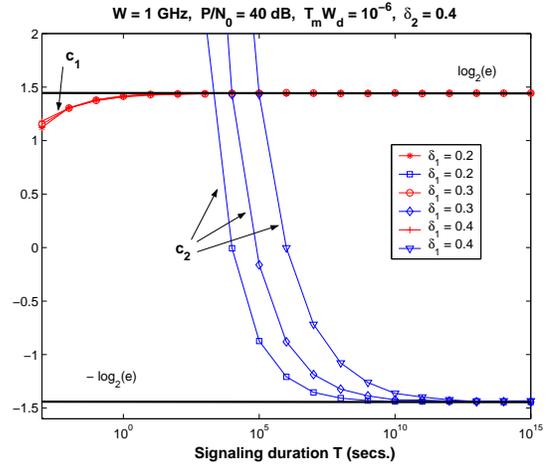} %EbNo.eps}
\caption{\sl Convergence of the coefficients of the $\snr$ and
$\snr^{2}$ terms in capacity as a function of $T$.}
\label{fig_c1c2}
\end{center}
\end{figure}

Note that $\mu_{\eff} \rightarrow \infty$ as $T \rightarrow
\infty$ for sparse channels, which implies that first- and
second-order optimality can be achieved by simply increasing $T$.
This is due to the impact of sparsity in Doppler and in direct
contrast to the case of rich multipath where the coherence
requirement is independent of signaling duration. We provide
numerical illustration of the results by considering the low
$\snr$ asymptote of the coherent capacity in (\ref{coh_bounds}).
The coefficients of the first- and second-order terms are
$\lambda_{1} = \log_{2}(e)$ and $\lambda_{2} = - \log_{2}(e)$,
respectively. In Fig.~\ref{fig_c1c2}, we plot the numerically
estimated values $c_1$ and $c_2$ of $\lambda_1$ and $\lambda_2$,
respectively, for the training-based communication scheme, which
are estimated using Monte-Carlo simulations. We observe that for a
large enough $T$ such that $\mu_{eff} > 3$, the second-order
constant $c_{2} \rightarrow \lambda_{2} = - \log_{2}(e)$. Also
shown in the figure is the behavior of the first-order constant
and it is seen that $c_{1} \rightarrow \lambda_{1}$ for a much
smaller value of $T$ since all we need is $\mu > 1$.

Contrary to the traditional emphasis on peaky signaling to improve
the spectral efficiency of non-coherent communication, our results
imply that delay-Doppler sparsity, along with a suitable choice of
$T$, $W$ and $P$ as in (\ref{TvsW}) is sufficient to achieve a
desired level of coherence with \emph{non-peaky} signaling
schemes. It is shown in \cite{capacity_sparse_STSP} that the
$\Tcoh$ requirements for non-peaky signals under our framework are
still better than those for peaky training-based communication
schemes proposed in \cite{zheng} based on a rich multipath
assumption.

%% file: reliability.tex
\section{Reliability of Sparse Multipath Channels}
\label{sec4} The reliability function of the channel is defined as
\cite{gallager_book}
\begin{equation*}
E \left( R \right) = \lim_{N \rightarrow \infty} \text{sup}
\frac{-\log \: \pe \left(N,R \right)}{N}
\end{equation*}
where $\pe \left(N,R \right)$ is the average probability of error
over an ensemble of codes (random coding) in which each codeword
spans the signal space dimensions $N = TW$ and communication takes
place at transmission rate $R$. For any finite $N$, while the
random coding exponent $\er (N,R)$ provides a lower bound to
$E(R)$, the sphere-packing exponent $E_{sp} (N,R)$ is an upper
bound to $E(R)$ \cite{gallager_book}.
\begin{figure*}[]
\begin{eqnarray}
\er^{tr} \left(N,R\right) & = &  \begin{cases} \frac{1}{\nc} \log
\left(1+ \frac{\left(\nc - 1 \right) K^{*} \left(1 - (K^{*})^{1-\epsilon} \right)}{2} \right)- R - \littleo(1)  & 0 \leq R \leq \rcr \\
\frac{1}{\nc} \log \left(1+ \frac{\left(\nc - 1 \right) K^{*} \left(1 - (K^{*})^{1-\epsilon} \right) \rho^{*}}{1+\rho^{*}} \right) - \rho^{*}R - \littleo(1) & \rcr \leq R \leq \rmax \\
0 & R > \rmax \end{cases} \label{err_exp_tr} \\
\rcr & = & \tsty{ \left(\frac{\nc-1}{2\nc} \right)
\frac{K^{*}\left(1 - (K^{*})^{1-\epsilon} \right)}{\left[2+
\left(\nc-1\right) K^{*}\left(1 - (K^{*})^{1-\epsilon} \right)
\right]} } \label{rcrtr} \\
\rmax & = & \tsty{ \left(\frac{\nc-1}{\nc} \right) K^{*}\left(1 - (K^{*})^{1-\epsilon} \right) }  \label{rmaxtr} \\
K^{*} & = & \tsty{ \frac{\etas \left(1-\etas \right) \left( \nsnr
\right)^{2} }{\left(\nc-1\right)\left(1+ \etas
\nsnr \right) + \left(1-\etas \right) \nsnr} } \label{snrp} \\
\etas & = & \tsty{ \frac{\nsnr+\nc-1}{(\nc-2)\nsnr} \left[\sqrt{1+
\frac{(\nc-2)\nsnr}{\nsnr+\nc-1}} - 1 \right] } \label{etaopt} \\
\rho^{*} & = & \frac{-(2+k_{1}) + \sqrt{(2+k_{1})^{2} + 4
(1+k_{1})(\frac{k_{1}}{\nc R} - 1)}}{2 (1+k_{1})} \hsp \,\ \hsp
\hsp \hsp k_{1} = (\nc-1) K^{*} \left(1 - (K^{*})^{1-\epsilon}
\right) \label{rhoopt}
\end{eqnarray}
\hrulefill %\vspace*{2pt}
\end{figure*}
We recall the random coding upper bound on $\pe$
\cite{gallager_book} given by
\begin{eqnarray}
\pe & \leq &  e^{-N \left[\er \left(N,R \right) \right]}
\label{perr} \\
\er \left(N,R \right) &  = & \max_{0 \leq \rho \leq 1} \max_{Q}
\left[ \eo \left(N,\rho, Q \right) - \rho R \right]
\label{randexp}
\end{eqnarray}
\begin{equation}
\begin{split}
&\tsty{\eo \left(N,\rho, Q \right)} = \\ & \tsty{ -\frac{1}{N}}
\tsty{ \log \left( E_{\mathbf{H}} \left[ \int_{\by} \left[
\int_{\bx} q(\bx) p(\by | \bx,\bH)^{\frac{1}{1+\rho}} dx
\right]^{1+\rho} dy \right] \right) } \label{eo}
\end{split}
\end{equation}
We compute the random coding error exponent in (\ref{randexp}) for
the training and communication scheme described in
Sec.~\ref{sec2c}. The result is summarized in the following
theorem.
\begin{thm}
The average probability of error for the training-based
communication scheme is upper-bounded by,
\begin{equation}
\pe \leq e^{-N \left[ \er^{tr} \left(N,R\right) \right]} \nonumber
\end{equation}
where $\er^{tr} \left(N,R\right)$ is given in (\ref{err_exp_tr})
on the next page. $\rmax$ in (\ref{rmaxtr}) defines the maximum
rate until which we have a non-zero error exponent (decaying
$\pe$). The critical rate, $\rcr$ in (\ref{rcrtr}) delineates the
regime of the optimal parameter $\rho^{*}$ that maximizes the
exponent. $\rho^{*} = 1$ for $0 < R < \rcr$ and $\rho^{*}$ is
given in (\ref{rhoopt}) for $\rcr < R < \rmax$. The constant
$\epsilon > 0$ and is chosen very small ($\epsilon \rightarrow 0$)
so that the $\littleo(1)$ terms are negligible. See
\cite{allerton_06} for more details.
\label{thm2}
\end{thm}
Note that the error exponent of the training and communication
scheme in (\ref{err_exp_tr}) depends \emph{only} on $\snr$ and
$\nc = \frac{k}{\snr^{\mu}}$.

\subsection{Discussion of Results on Reliability}
\label{sec4a} 
\begin{figure}[width=2.8in]
\begin{center}
\includegraphics[width=2.8in]{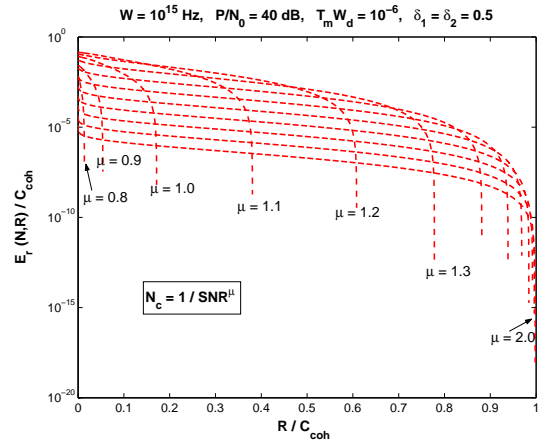} %EbNo.eps}
\caption{\sl Random coding error exponent versus rate for a sparse
channel and for the training-based communication scheme. Different
curves correspond to different values of $\mu$ in the key
relationship $\nc = \frac{1}{\snr^{\mu}}$.} \label{fig_err_exp}
\end{center}
\end{figure}

We investigate the behavior of the random coding
exponent for different values of $\mu$ as illustrated in
Fig.~\ref{fig_err_exp} for the given channel parameter set. It is
observed that for any transmission rate $R$, there exists an
optimum value of $\mu = \mu_{opt}(R)$ for which the error exponent
in (\ref{err_exp_tr}) is maximum. For any $N$, we formally define
\begin{equation}
\mu_{opt} (N,R) = \arg \max \limits_{\mu} \left[ \er \left(R, N,
\mu \right) \right] \label{muopt}
\end{equation}
where we have written $ \er \left(R, N, \mu \right) = \left[
\er^{tr} \left(N,R\right) \right]$ in (\ref{err_exp_tr})
explicitly as a function of $\mu$ to emphasize its dependance. As
we traverse from $R = 0$ to $R = C_{coh}$ (as in
(\ref{exact_cap})), the optimal operating point at each rate is
dictated by the value of $\mu_{opt}$ in (\ref{muopt}) and can be
\emph{achieved} by choosing $T$, $W$ and $P$ as in
(\ref{TWP_locus}). Furthermore, the optimizing $\mu_{opt}$
increases monotonically as we consider larger transmission rates.
In fact, using the results on capacity from Theorem~\ref{thm1}, it
follows that with $\mu = 1$, we only obtain first-order optimality
and therefore the error exponent in Fig.~\ref{fig_err_exp} is
non-zero only for a fraction of the coherent capacity, $C_{coh}$.
On the other hand, at $R = C_{coh}$ , we would require $\mu_{opt}
> 3$ (second-order optimal) in order to achieve a positive error
exponent.

\begin{figure}[width=2.8in]
\begin{center}
\includegraphics[width=2.8in]{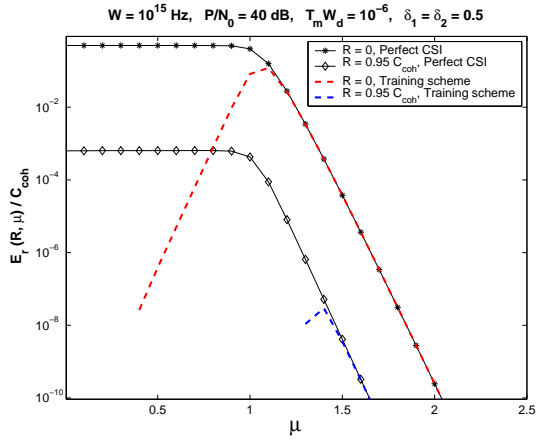} %EbNo.eps}
\caption{\sl Illustration of the learnability versus diversity
tradeoff for sparse channels. The value of $\mu$ at which the
maximum is attained in each case defines $\mu_{opt}(R)$ at the
corresponding transmission rate $R$.} \label{fig_err_exp_vsmu}
\end{center}
\end{figure}

In Fig.~\ref{fig_err_exp_vsmu}, we plot the error exponent in
(\ref{err_exp_tr}) as a function of the parameter $\mu$ for two
different transmission rates. For each scenario, we observe that
the error exponent is concave as a function of $\mu$ and is
maximized at $\mu = \mu_{opt}$. Also illustrated in the figure is
the error exponent with perfect CSI at the receiver that is an
upper bound to $\er^{tr}(N,R)$ in (\ref{err_exp_tr}) and decreases
monotonically with $\mu$. These plots reveal a fundamental
\emph{learnability versus diversity tradeoff} in sparse channels.
For any rate $R$, when $\mu < \mu_{opt}(R)$ (too little
coherence), the system is in the learnability-limited regime and
the error exponent of the training-based communication scheme is
smaller due to poor channel estimation performance. On the other
hand when $\mu > \mu_{opt}(R)$ (too much coherence), we are in the
diversity-limited regime and the hit taken by the error exponent
here is due to the inherent reduction in the degrees of freedom
(DoF) (or delay-Doppler diversity, $D$). The best exponent is
obtained at $\mu = \mu_{opt}(R)$, which demarcates the two regimes
and describes the optimal tradeoff.

